\documentclass[letter]{aa}  

\usepackage{color}
\usepackage{xcolor}
\usepackage{txfonts}
\usepackage[version=4]{mhchem}
\usepackage{placeins}
\usepackage{multirow}
\usepackage{verbatim}

\begin{document}
\title{The dry and carbon poor inner disk of TW Hya:\\evidence for a massive icy dust trap }
\author{Arthur D. Bosman \inst{1}\thanks{\emph{Present address:} Department of Astronomy, University of Michigan, 311 West Hall, 1085 S. University Avenue, Ann Arbor, MI 48109, USA} \and Andrea Banzatti \inst{2} }
\institute{
Leiden Observatory, Leiden University, PO Box 9513, 2300 RA Leiden, The Netherlands \\  \email{arbos@umich.edu}
\and
Department of Physics, Texas State University, 749 N Comanche Street, San Marcos, TX 78666, USA }
  \abstract
   {Gas giants accrete their envelopes from the gas and dust of proto-planetary disks, so it is important to determine the composition of the inner few AU, where most giant planets are expected to form. }
   {We aim to constrain the elemental carbon and oxygen abundance in the inner disk ($R<$2.3 AU) of TW Hya and compare with the outer disk ($R>2.3$ AU) where carbon and oxygen appear underabundant by a factor of $\sim$50. }
   {Archival \textit{Spitzer}-IRS and VLT-CRIRES observations of TW Hya are compared with a detailed thermo-chemical model, DALI. The inner disk gas mass and elemental C and O abundances are varied to fit the mid-infrared \ce{H2} and \ce{H2O} line fluxes as well as the near-infrared CO line flux.}
   {Best fitting models have an inner disk that has a gas mass of $ 2 \times 10^{-4} M_\odot$ with $\ce{C}/\ce{H} \approx 3 \times 10^{-6}$ and $\ce{O}/\ce{H} \approx 6 \times 10^{-6}$. The elemental oxygen and carbon abundances of the inner disk are $\sim 50$ times underabundant compared to the ISM and are consistent with those found in the outer disk.}
   {The uniformly low volatile abundances imply that the inner disk is not enriched by ices on drifting bodies that evaporate. This indicates that drifting grains are stopped in a dust trap outside the water ice line. Such a dust trap would also form a cavity as seen in high resolution sub-millimeter continuum observations. If CO is the major carbon carrier in the ices, dust needs to be trapped efficiently outside the CO ice line of $\sim$20 AU. This would imply that the shallow sub-millimeter rings in the TW Hya disk outside of 20 AU correspond to very efficient dust traps. The more likely scenario is that more than 98\% of the CO has been converted into less volatile species, e.g. \ce{CO2} and \ce{CH3OH}. A giant planet forming in the inner disk would be accreting gas with low carbon and oxygen abundances as well as very little icy dust, potentially leading to a planet atmosphere with strongly substellar C/H and O/H ratios. }

\titlerunning{The dry and carbon poor inner disk of TW Hya}
\maketitle

\section{Introduction}

The elemental abundances of carbon and oxygen in proto-planetary disks are a vital input to planet formation models and, combined with characterization of exoplanets, can tell us about the formation history of those planets. The simple picture of elemental abundances changing statically at the icelines of the main chemical species \citep[e.g.][]{Oberg2011} is now being growingly enriched by a number of observational and modeling studies. Chemical evolution can be efficient in changing the composition of the disk gas, changing the major carriers of carbon and oxygen \citep[e.g.][]{Eistrup2016, Schwarz2018, Bosman2018}. Furthermore, disk dynamics and dust evolution can efficiently transport the volatile component of the disk, changing the elemental composition of the gas and ice \citep{Kama2016, Booth2017, Bosman2018, Krijt2016Water, Krijt2018}. 

As the interplay of the physical and chemical processes is complex, observations are needed to benchmark disk models and provide the much needed input for planet formation models. \ce{CO} and \ce{H2O} observations focusing on the outer regions of proto-planetary disks show that their abundances are up to two orders of magnitude lower than expected, implying that chemical and physical processes are indeed modifying the abundances of these species \citep{Hogerheijde2011, Favre2013, Bergin2013, Miotello2017, Du2017}. Planet formation models, however, primarily need the composition around and within the \ce{H2O} ice line, as this is where we expect giant planets to form and accrete their atmospheres \citep[e.g.][]{Kennedy2008,Cridland2017,Dawson2018}.

Therefore we aim to constrain the elemental abundances in the inner disk of TW Hya. The outer disk of TW Hya has been well studied and the elemental composition of the gas has been constrained from a variety of observations, including HD to trace the total disk mass, allowing for the measurement of absolute abundances \citep{Hogerheijde2011, Bergin2013, Kama2016, Schwarz2016, Trapman2017}. These efforts have shown that both the volatile carbon and oxygen abundance in the outer disk are lowered by a factor of $\sim$50 compared to the ISM. Observations of \ce{^{13}C^{18}O} in TW Hya show that, while some CO comes of the grains within the \ce{CO} ice line, the CO abundance stays a factor of $\sim$20 lower than the ISM abundance \citep{Zhang2017}, implying that a lot of the carbon is trapped on the grains or converted into other species.  

The physical and chemical structure of the inner disk of TW Hya has also been studied in detail. It is the disk that has been resolved at the highest physical resolution with ALMA \citep{Andrews2016}. Together with abundant photometry and infrared interferometry, there is a good picture of the inner disk structure of TW Hya \citep{Andrews2012, Menu2014, Kama2016}. In the infrared, high signal-to-noise, \textit{Spitzer}-IRS and VLT-CRIRES spectra have been taken. These observations include 3.2 km s$^{-1}$ resolution observations of the CO $v=1-0$ rovibrational band at 4.7 $\mu$m, and detections of the \ce{CO2} 15$\mu$m $Q$-branch, pure rotational lines lines of \ce{H2}, \ce{H2O} and \ce{OH}, as well as a number of atomic hydrogen lines \citep{Najita2010,Pontoppidan2008spectro}. Modeling efforts have further constrained the inner disk gas mass using \ce{H2} \citep{Gorti2011} as well as the \ce{H2O} content of the inner disk \citep{Zhang2013}. These studies hint at abundances for \ce{CO} and \ce{H2O} in the inner disk that are lower than expected from inner disk chemistry for gas of ISM composition, implying that elemental carbon and oxygen are depleted relative to the ISM.

Here we will build on these studies, using the better constrained outer disk structure and composition from \citet{Trapman2017}, with an updated inner disk gap from ALMA observations \citep{Andrews2016}. Using this disk structure and the thermo-chemical code DALI we constrain the elemental carbon and oxygen abundance in the inner disk of TW Hya. 

\section{Methods}

The lines and features and their fluxes that we consider for our modeling comparison are tabulated in Table~\ref{tab:obsdat}. The CO flux for TW Hya is taken from \citet{Banzatti2017} based on VLT-CRIRES \citep{Kaufl2004} spectra presented in \citet{Pontoppidan2008spectro} (program ID 179.C-0151). The fluxes for \ce{H2} and \ce{H2O} are extracted from the \textit{Spitzer}-IRS spectrum obtained from program GO 30300 using the method described in \citet{Banzatti2012}. The spectrum has been published in \citet{Najita2010} and \citet{Zhang2013}.

\begin{table*}[t]
\centering
\caption{\label{tab:obsdat} \ce{H2}, \ce{CO} and \ce{H2O} fluxes from observations and modeling}
\begin{tabular}{l c c c c c }

\hline
\hline
Molecular & Transition\tablefootmark{1} & wavelength ($\mu$m) & Observed flux\tablefootmark{2} & Model flux\tablefootmark{2} & $E_\mathrm{up}$ (K) \\
 \hline
\ce{H2} & $J=3-1$ $(S(1))$  & 17.0 & 1.57$\pm0.05$ & $1.7$ & 1015\\
\ce{H2} & $J=4-2$ $(S(2))$ & 12.3 & 1.21$\pm0.20$ & $0.75$ & 1681\\
\ce{CO} & $v,J=1,10-0,11$ ($v1$ $P(10)$) & 4.75 & $0.64 \pm 0.05$ & $0.72$ & 3330 \\
\ce{H2O} & $12_{5,8}$ -- $11_{2,9}$ & 17.12 & $< 0.17$ & $0.03$ & 3243\\
\ce{H2O} & $11_{3,9}$ -- $10_{0,10}$ & 17.22 & $< 0.29$ & $0.05$ & 2473\\
\ce{H2O} & $11_{2,9}$ -- $10_{1,10}$ & 17.36 & $< 0.32$ & $0.11$ & 2396\\
\ce{H2O} & $9_{2,7}$ -- $8_{1,8}$ & 21.85 & $2.8 \pm 0.2$ & $6.7$ & 1730 \\ 
\ce{H2O} & $9_{4,5}$ -- $8_{3,6}$ & 28.23 & $2.1 \pm 0.1$ & $1.3$ & 1929 \\
         & $9_{5,4}$ -- $8_{4,5}$ &       &               &       & 2122 \\
\ce{H2O} & $8_{6,3}$ -- $7_{5,2}$ & 28.43 & $1.6 \pm 0.2$ & $1.2$ & 2031\\ 
& $8_{6,2}$ -- $7_{5,3}$ &       &               &       & 2031 \\
\ce{H2O} & $7_{7,0}$ -- $6_{6,1}$ & 28.6 & $1.7 \pm 0.1$ & $1.2$ & 2006\\ 
& $7_{7,1}$ -- $6_{6,0}$ &       &               &       & 2006 \\
\ce{H2O} & $6_{3,1}$ -- $5_{0,5}$ & 30.87 & $2.7 \pm 0.2$ & $1.8$ & 934 \\ 
&$8_{5,4}$ -- $7_{4,3}$ &       &               &       & 1806 \\
\ce{H2O} & $6_{6,1}$ -- $5_{5,0}$ & 33   & $3.96 \pm 0.46$ & $4.0$ & 1504\\
&$7_{5,2}$ -- $6_{4,3}$&       &               &       & 1524 \\ 
\hline
\end{tabular}
\tablefoot{Observed \ce{CO} line flux is taken from \citet{Banzatti2017}, \ce{H2} and \ce{H2O} fluxes are extracted from the \textit{Spitzer}-IRS spectra using the method in \citet{Banzatti2012}. The table denotes the upper level energy of the dominant, according to our model, transitions of each blend. The \ce{H2} $S(2)$ line is contaminated by an OH line; the contribution of the OH line was estimated by taking the average flux of two adjacent OH lines. }
\raggedright
\tablefoottext{1. In the case of \ce{H2O}, the dominant transition(s) of the feature, 2. In units of $10^{-14}$ erg s$^{-1}$ cm$^{-2}$}

\end{table*}

\begin{figure}[t]
\centering
\includegraphics[width = \hsize]{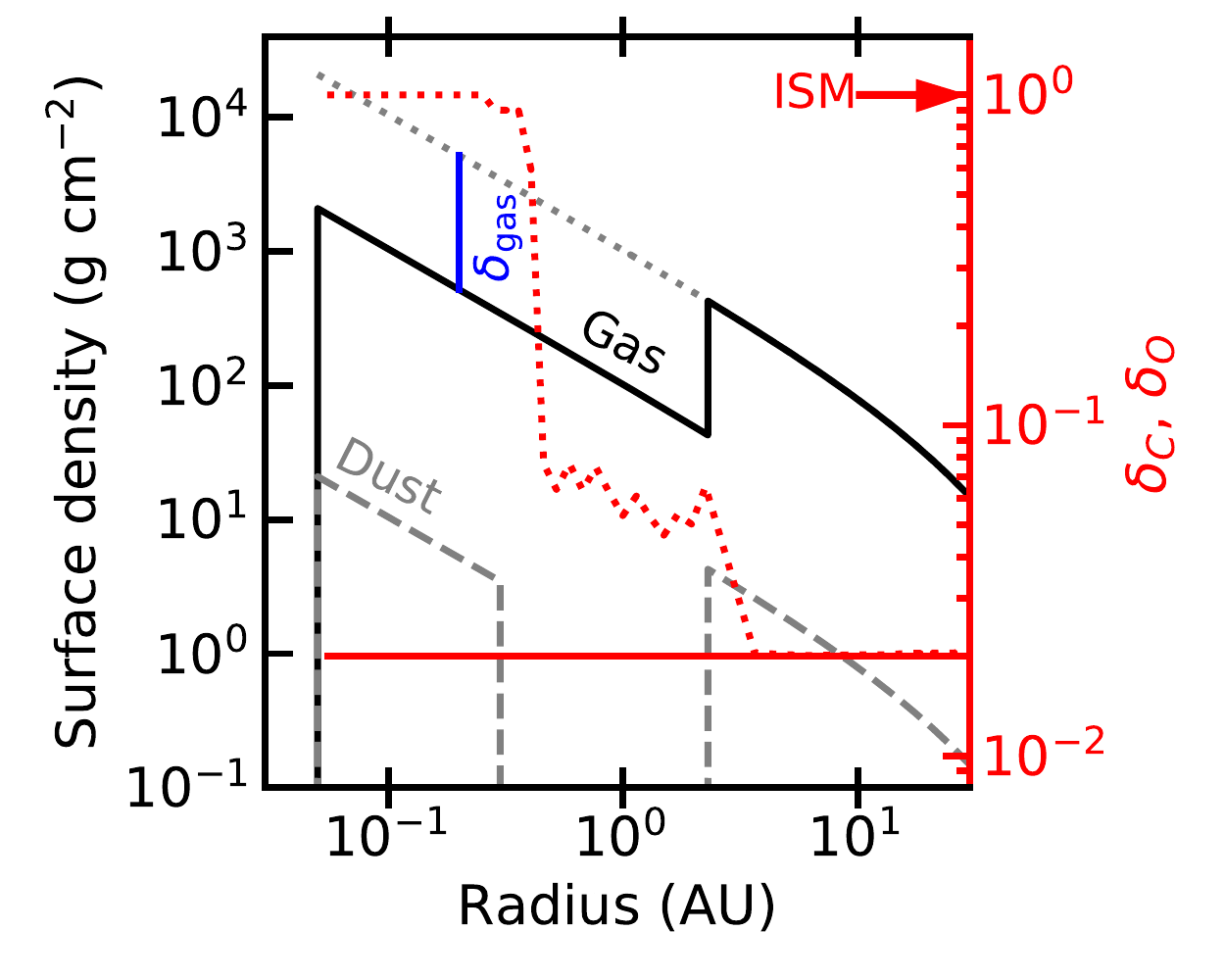}
\caption{\label{fig:surf_dens} Surface density structure in inner regions of the TW Hya model. $\delta_\mathrm{gas}$ is the inner disk drop in gas surface density. The value $\delta_\mathrm{gas} = 0.1$ used here best fits the \ce{H2} observations. The red curves show the column averaged oxygen and carbon depletion factors in TW Hya. The red solid line shows a model with constant depletion. The dotted line shows the depletion profile assuming that carbon and oxygen return to the ISM values above $T_\mathrm{gas} = 150$ K.  } 
\end{figure}

\begin{figure}[t]
\includegraphics[width=\hsize]{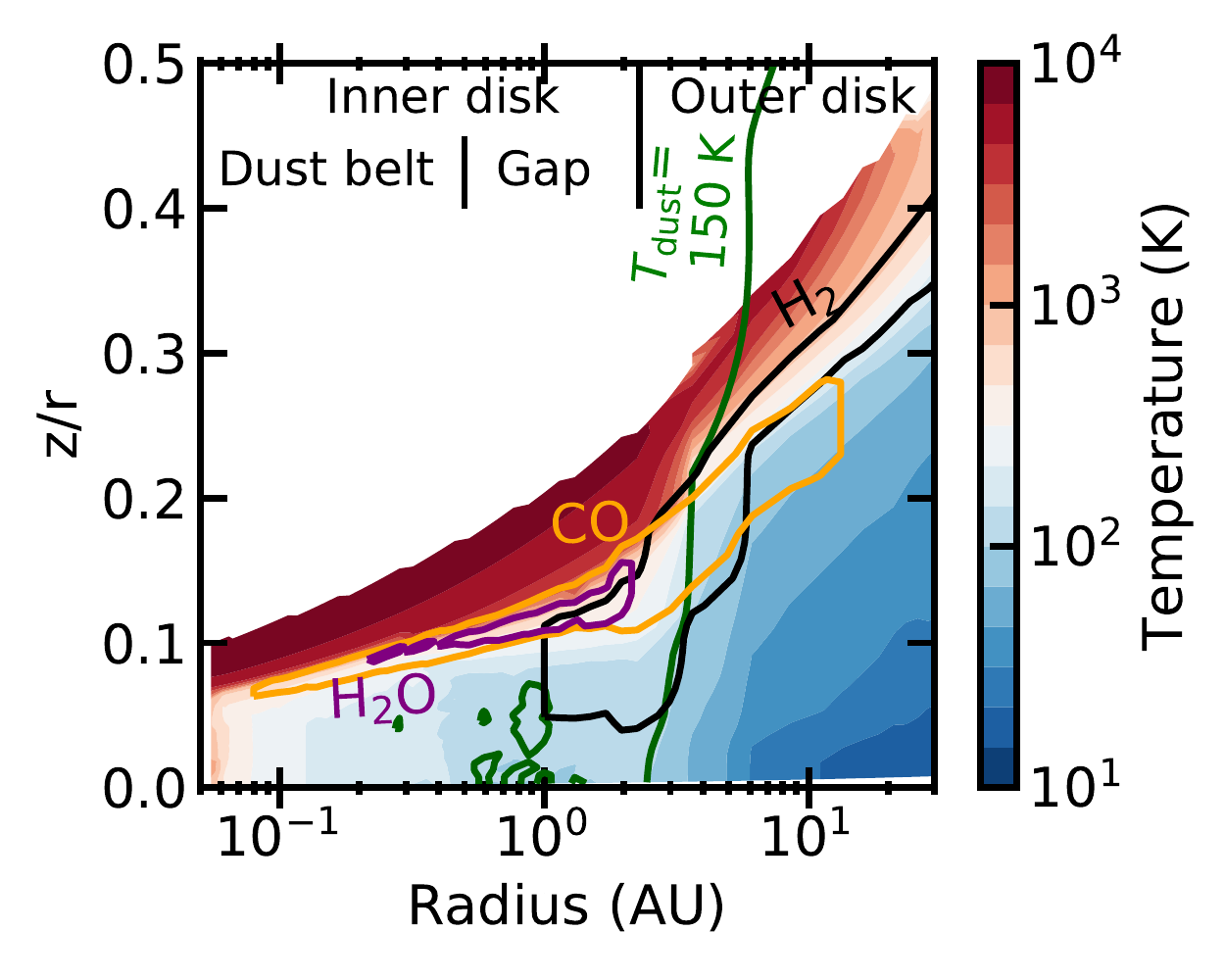}
\caption{\label{fig:Emitting areas} Map of the gas temperature in the inner region of the model together with the emitting areas of the \ce{H2} $S(1)$ line (black) the \ce{CO} $v1$ vibrational lines (orange) and the strongest line of the \ce{H2O} 33 $\mu$m feature (purple). The green line shows $T_\mathrm{dust} = 150$ K, the approximate location of the \ce{H2O} ice line. The emitting areas radially overlap in the inner disk. }
\end{figure}

We start with the Dust and Lines (DALI) \citep{Bruderer2012,Bruderer2013} model of \citet{Trapman2017} for TW Hya \citep[see also ][model parameters are given in Table~\ref{tab:model}]{Kama2016}. This model fits the SED, many far-infrared and sub-millimeter lines as well as the ALMA \ce{^{12}CO} $J=3-2$ image. The most important lines are the HD 112 and 56 $\mu$m lines, constraining the outer disk mass, and many CO (isotopologue) lines constraining the carbon abundance. On top of these some atomic carbon and oxygen fine-structure lines have also been fit, which constrain the elemental carbon and oxygen abundances in the outer disk at C/H = $2.7 \times 10^{-6}$ and O/H = $5.8 \times 10^{-6}$ a factor 50 lower than expected for the interstellar medium (ISM). In this work we update the inner disk structure moving the outer edge of the gap from 4 to 2.4 AU in accordance with the bright sub-millimeter ring seen by \citet{Andrews2016} at this radius (assuming a distance of 54 parsec). 

\begin{figure*}[t]
\includegraphics[width=\hsize]{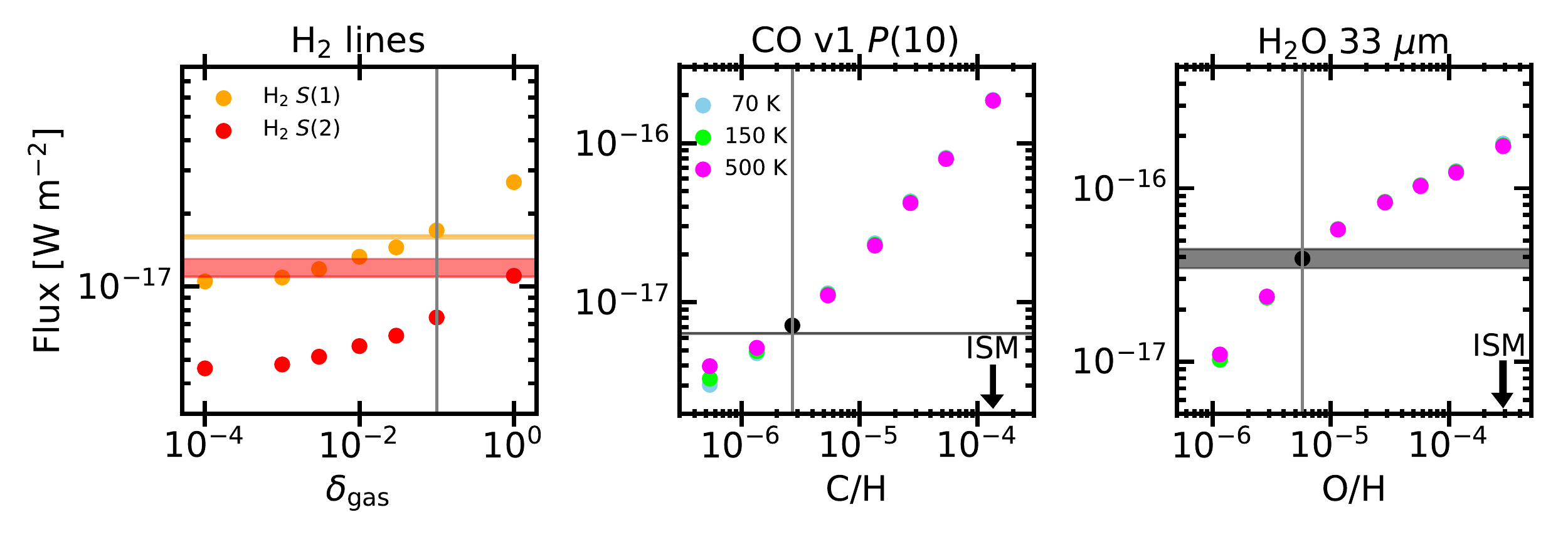}
\caption{\label{fig:fluxes_all} Comparison of the \ce{H2} line fluxes  (\textit{left}), CO rovibrational line flux(\textit{middle}) and the \ce{H2O} 33 $\mu$m feature flux (\textit{right}) between models and data. For the left panel the amount of gas within 2.4 AU is varied. For the \textit{middle} and \textit{right} panels $\delta_\mathrm{gas} = 0.1$ is used and C/H and O/H are varied respectively. Horizontal bands show the 1 $\sigma$ variation of the observed fluxes. The different coloured points denote different values for $T_\mathrm{step}$. The temperature threshold above which the abundances are varied. The vertical lines in the middle and right panel show the volatile carbon and oxygen abundances assumed for the ISM \citep{Meyer1998, Lacy2017}.}
\end{figure*}

Figure 1 shows the surface density structure in the inner disk.
The inner disk mass is varied by varying $\delta_{\mathrm{gas}}$ in the inner disk using a model that has constant, low elemental abundances. The predicted \ce{H2} lines for DALI are compared to the observed fluxes. The model that fits best is then used to constrain the elemental C and O abundances. We make the simplifying assumptions that the CO abundance in the inner disk scales linearly with the total elemental carbon abundance, that the \ce{H2O} abundance scales with the elemental oxygen that is not locked in CO and that changing the CO and \ce{H2O} abundances do not significantly alter the gas temperatures. The abundances of CO and \ce{H2O} multiplied by a factor $\delta_C$ and $\delta_O$, respectively, above gas temperatures of $T_\mathrm{step}$ of 70, 150 and 500 K. The 70 and 150 K are the sublimation temperatures of \ce{CO2} and \ce{H2O} respectively, whereas 500 K as a rough transition temperature for the release of carbon from a more refractory reservoir. The excitation is recalculated for these new abundance structures and the \ce{CO} v1 P(10) line flux as well as the full \ce{H2O} spectrum between 12 and 34 $\mu$m are extracted.

The excitation for \ce{H2} is done in LTE as there are no collisional rate coefficients available for \ce{H2}, however, the low Einstein A coefficients of the \ce{H2} lines mean that deviations from LTE are expected to be negligible \citep{Wolniewicz1998}. For \ce{H2O} and \ce{CO} the local excitation, de-excitation balance is calculated explicitly. For CO, collision rate coefficients from \citet{Yang2010} for \ce{H2} and \citet{Song2015,Walker2015} for \ce{H} are used \citep[see also][]{Bosman2019}. For \ce{H2O}, the data file from the LAMDA database\footnote{\url{http://strw.leidenuniv.nl/~moldata}} \citep{Schoier2005} is used. 

\section{Results}

Figure~\ref{fig:Emitting areas} compares the emitting regions of the different species in the constant abundance model with $\delta_\mathrm{gas} = 0.1$. There is a large overlap in the emitting areas, especially in the inner disk ($<2.4$ AU). \ce{H2O} has the most confined emitting area and only probes the inner disk while \ce{CO} and \ce{H2} both also probe the outer disk, where the composition of the gas is already strongly constrained \citep{Bergin2013, Kama2016, Trapman2017, Zhang2017}.

Figure~\ref{fig:fluxes_all} compares the results of the DALI modeling with the observed fluxes. The \ce{H2} fluxes constrain the inner disk gas mass to be around $1.7 \times 10^{-4} M_\odot$ ($\delta_\mathrm{gas} \approx 0.1$), comparable to the value found in \citep{Gorti2011}. The observations show a higher \ce{H2} S(2)/S(1) line ratio compared to the models, this indicates that the average gas temperature in the inner disk of the model is too low. The upper level energies of the CO ($\sim$3000 K) and \ce{H2O} ($\sim$1500 K) are similar or higher than the upper level energies of the \ce{H2} lines ($\sim$1000 and $\sim$1700 K), therefore a higher temperature in the inner disk would lead to even lower inferred values for C/H and O/H.

Both the \ce{CO} lines and \ce{H2O} lines fit well to an inner disk elemental abundance that is similar to abundance found in the outer disk, that is, a factor 50 depleted compared to the ISM. Even a jump of a factor of 2 in the elemental carbon or oxygen abundance above 500 K can be ruled out based on the observed fluxes. The \ce{H2O} fluxes in Table~\ref{tab:obsdat} show that this model underpredicts the \ce{H2O} detections between 20 and 31 $\mu$m by a factor $\sim$2, this is in line with the gas temperature in the model being too low.

%
%
%

\section{Discussion}

\subsection{Constraining the inner disk chemical structure}

Here, using a more complete physical and chemical structure of both the inner and the outer disk, we can quantify the total \ce{H2} mass, as well as the volatile carbon and oxygen abundance in the inner disk, confirming that the inner disk of TW Hya is both oxygen and carbon poor by a factor $\sim$ 50. Furthermore, our modeling shows that there is no significant (factor 2 or more) increase in volatile carbon or oxygen in the inner disk. Thus, there is no sign of volatile release at the \ce{CO2} or \ce{H2O} ice lines nor is there evidence of carbonaceous or silicate grain destruction at $T < 500$ K.  

Both \citet{Gorti2011} and \citet{Zhang2013} have studied the inner region of TW Hya using detailed modeling. They note that they overproduce the CO rovibrational lines by a factor of $\sim 2$. Furthermore, \citet{Gorti2011} assume LTE excitation for CO, which generally underpredicts fluxes compared to models that include infrared pumping of the vibrational levels by the inner disk continuum emission \citep{Bruderer2015, Bosman2017}. These two factors together explain why our models need a CO or elemental carbon abundance that is almost two orders lower than ISM to fit the CO rovibrational line. 

\citet{Zhang2013} use a low abundance in the inner disk $x_\mathrm{H2O} < 10^{-6}$ with a ring of high abundance \ce{H2O} at 4 AU, their ice line, to produce the \textit{Spitzer} lines. Including such a ring and fitting the 33 $\mu$m ($E_\mathrm{up} = 1504$ K) feature would lead to inferring lower oxygen abundances in the inner disk, however, it would also move the \ce{H2O} emission to colder gas, increasing the discrepancy between model and observations for the 22 and 28 $\mu$m features. A constant abundance model, thus fits the data better than a ring model.

\subsection{Hiding C and O carriers?}
The DALI models predict that CO and \ce{H2O} are the dominant gas-phase oxygen and carbon carriers in the inner disk. However, it is possible that carbon and oxygen are locked in other gaseous species.
There are a few obvious candidates for hiding more carbon and oxygen in molecular gas: \ce{CO2}, \ce{C2H2}, \ce{HCN} and  \ce{CH4}. There is a \ce{CO2} feature detected in the \textit{Spitzer} spectrum of TW Hya. Applying the model results from \citet{Bosman2017} to the detected feature, retrieves a \ce{CO2} abundance lower than $10^{-9}$ w.r.t. \ce{H2}. The models in \citet{Bosman2017} do not have a gap, which, if included would only increase the strength of the \ce{CO2} lines. As such the \ce{CO2} abundance $<10^{-9}$ is a stringent upper limit and it is not an abundant carrier of either carbon or oxygen.

\ce{C2H2} and \ce{HCN} are detected in many proto-planetary disks \citep{Salyk2011}, however, neither are detected in the spectrum of TW Hya. As the $Q$-branches of \ce{HCN} and \ce{C2H2} around 14 $\mu$m have similar upper level energies and Einstein A coefficients to the 15 $\mu$m Q-branch of \ce{CO2}, both \ce{HCN} and \ce{C2H2} should have been detected if they are as abundant as \ce{CO2}. The lack of observed features implies that both \ce{HCN} and \ce{C2H2} contain less than 1\% of the volatile carbon. The final possible gaseous carbon carrier is \ce{CH4}. Observations of this molecule have so far only been successful in one disk, in absorption thus the \ce{CH4} abundance is not strongly constrained by observations \citep{Gibb2013}, however, from chemical models is not expected to be the most abundant carbon carrier in inner disk atmospheres \citep{Walsh2015, Agundez2018}. As such, with \ce{CO} and \ce{H2O} we trace the bulk of gaseous carbon and oxygen in the inner disk.

\subsection{Implications of uniform depletion}
\begin{figure}
\centering
\includegraphics[width = \hsize]{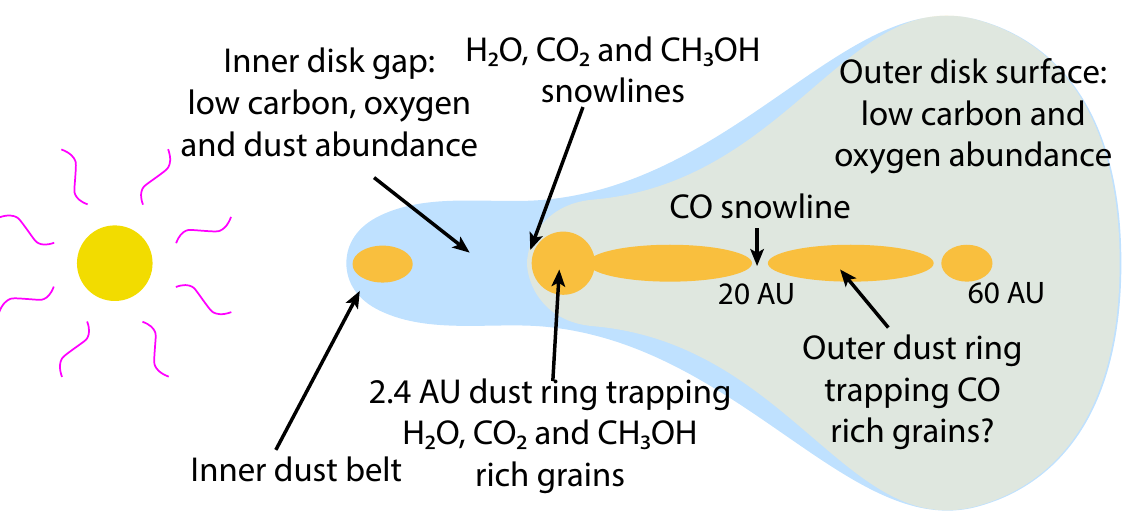}
\caption{\label{fig:schem} Schematic of the TW Hya disk showing the locations that should trap the oxygen and carbon bearing molecules. Oxygen needs to be trapped outside of 2.4 AU and is most likely trapped in the dust ring at that location. Carbon can either be trapped at the same location, with \ce{CO2} as most probable carrier, or at larger radii outside the CO ice line, with CO as the most probable carrier.}
\end{figure}

The low elemental abundances in the inner disk imply one of two things: either the entire disk is depleted in volatiles, implying dust grains with almost no ices, or volatiles are efficiently locked in ices on solids that are not transported through the disk. 

To deplete the entire disk in volatiles, a period of very strong radial drift, coupled with strong radial mixing is necessary \citep[e.g.][]{Booth2017}. In such a scenario, the icy grains drift from the mass reservoir in the outer disk, into the regions of the disk where the ices desorb. This leads to enhancements in C/H and O/H in the inner regions of the disk. After a few Myr all of this high C/H and O/H gas is accreted onto the star and the disk is left depleted in carbon and oxygen. However, a disk like this will also be strongly depleted in solid material as the grains that transport the ices inward, will continue to drift into the star leading to dust depletions at least as high as the volatile element depletion. This scenario is thus very unlikely for TW Hya, which has a massive dust component, and a gas-to-dust ratio of around 100 \citep{Bergin2013}.

As efficient transport is excluded, this leaves the locking of 98\% of the carbon and oxygen in solids that are not efficiently transported. Assuming that CO and \ce{H2O} are the dominant oxygen and carbon carriers in the outer disk, it is necessary to stop \ce{CO} from being transported over the \ce{CO} ice line by a dust trap at a location larger than where the CO ice line is located. For TW Hya this would imply that one of the shallow millimeter continuum rings of TW Hya outside of 20 AU, corresponds to an efficient dust trap. An efficient dust trap at $>$ 20 AU should result in a disk with a large cavity. As dust is abundant down to 2.4 AU, a second dust trap would be necessary at that location, which would trap \ce{H2O} rich grains. 

In this scenario \ce{H2O} would be far more efficiently locked in large grains than \ce{CO} as the former is frozen out in a larger fraction of the disk. However, if gaseous \ce{CO} is efficiently converted into less volatiles species, especially \ce{CO2} and \ce{CH3OH}, then the CO and \ce{H2O} depletion fractions are likely to be more similar and an efficient dust trap around 20 AU is no longer strictly necessary. The age of TW Hya, 10 Myr, is long enough to convert large amounts of \ce{CO} into other species \citep{Donaldson2016, Bosman2018CO}. In this case, the dust should not be allowed to pass the \ce{CO2} or \ce{CH3OH} ice lines. These ice lines are at nearly the same location as the \ce{H2O} ice line, at the inner edge of the outer disk. As such, a dust trap at the innermost sub-millimeter ring would trap all of the icy \ce{CO2}, \ce{CH3OH} and \ce{H2O} in the outer disk. A schematic representation of this is given in Fig.~\ref{fig:schem}.


In summary, we find that the elemental carbon and oxygen abundance in the inner disk are lower by a factor $\sim$50 compared to the ISM. Even at temperatures of 500 K, the gaseous carbon or oxygen elemental abundances cannot have increased by a factor of 2, strongly constraining the release of volatiles and grain destruction up to 500 K. This is interpreted as the dust trap responsible for the dust free cavity also trapping the major carbon and oxygen bearing ices outside of 2.4 AU. A planet currently accreting gas in the gap will accrete very low amounts of carbon and oxygen, while possibly accreting ISM concentrations of nitrogen and noble gasses. Depending on the accretion history this planet could have a substellar C/H and O/H abundance.

%

\begin{acknowledgements}
Astrochemistry in Leiden is supported by the Netherlands Research School for
Astronomy (NOVA).

This work is partly based on observations made with CRIRES on ESO telescopes at the Paranal Observatory under program ID 179.C-0151. This work is based in part on observations made with the Spitzer Space Telescope, which is operated by the Jet Propulsion Laboratory, California Institute of Technology under a contract with NASA.
This project has made use of the SciPy stack \citep{Virtanen2019}, including NumPy
\citep{Oliphant2006} and Matplotlib \citep{Hunter2007}.
\end{acknowledgements}
\bibliographystyle{aa}
\bibliography{Lit_list}
\appendix

\section{DALI model}
\begin{table}[!h]
\centering
\caption{\label{tab:model}Parameters of the TW Hya model}
\begin{tabular}{l c}
\hline \hline
\multicolumn{2}{c}{Stellar parameters}\\
\hline
Stellar mass & 0.74 $M_\odot$\\ 
Stellar luminosity & 1 $L_\odot$\\ 
X-ray luminosity & $10^{30}$ erg s$^{-1}$ \\  
\hline
\multicolumn{2}{c}{Disk parameters}\\
\hline
Disk mass & 0.025 $M_\odot$ \\
Critical radius ($R_c$) & 35 AU \\
Surface density slope ($\gamma$) & 1\\
scale-height at $R_c$ ($h_c$) & 0.1 rad\\
Flaring angle ($\psi$) & 0.3 \\
Inner radius & 0.05 AU\\
Gap inner radius & 0.3 AU \\
Gap outer radius & 2.4 AU \\
Inner disk gas-to-dust & 100 \\
Inner disk gas depletion & [$10^{-5}$ -- 1] \\
Small dust size & 0.005--1 $\mu$m \\
Small dust fraction & 0.01\\
Large dust size & 0.005--1000 $\mu$m \\
Large dust fraction & 0.99\\
Large dust settling factor & 0.2 \\
\hline
\end{tabular}
\end{table}

\end{document}